# Frequency comb UV spectroscopy with one million resolved comb lines

ANDREY MURAVIEV,[1] DMITRII KONNOV,[1] SERGEY VASILYEV,[2] KONSTANTIN L. VODOPYANOV[1,*]

[1]CREOL, the College of Optics and Photonics, University of Central Florida, Orlando, Florida 32816, USA
[2] IPG Photonics Corporation, Marlborough, MA 01752, USA
*vodopyanov@creol.ucf.edu

**We perform dual-comb spectroscopy in two UV bands: 325-342 nm and 372-410 nm, corresponding to the 7th and 6th harmonics of a dual-comb 2.35-µm laser source, and simultaneously resolve ~500,000 and ~1,000,000 comb lines spaced by 80 MHz respectively.**

Dual-comb spectroscopy (DCS) has become a powerful technique, offering ultrabroad spectral coverage, high spectral resolution, and rapid data acquisition across the infrared to THz spectral range [1,2]. However, research on ultraviolet (UV) DCS is still in its early stages and remains relatively limited, which can be explained by the high phase and intensity noise of the existing UV combs and, at the same time, very tight (attosecond scale) constrains imposed on the time jitter between the interfering UV combs. UV DCS is highly desirable for its superior spectral resolution, providing unique insights into electronic transitions in atoms, molecules, and solids. Its applications are wide-ranging, from testing fundamental laws of physics to chemical analysis, photochemistry, trace gas sensing, and the search for exoplanets [3].

Xu et al. [4] achieved UV DCS by quadrupling the frequencies of 1550-nm electro-optic combs. Using photon-counting method, comb-line-resolved spectra near 389 nm (772 THz) were obtained with a span of 50 and 26 GHz at a line spacing of 500 and 200 MHz and a number of comb lines 100 and 130, respectively. A total coverage of 3 THz was achieved by gradually tuning the combs. By quadrupling the frequency of Yb-fiber (1030 nm) combs, McCauley et al. [5] demonstrated UV DCS between 260.5 and 263.5 nm with an instantaneous spectral span of 2 THz (0.5 nm) and resolution 1.2 GHz, providing 1,700 spectrally resolved data points. Fürst et al. [6] produced UV combs spanning 334-354 nm (spectral width 50.3 THz) via the 3rd harmonic of Yb-fiber combs. The spectral resolution of 50 GHz in this work corresponded to 1,000 resolved data points. In both of the above papers the authors used time-domain apodization, so that the comb lines were not resolved. Chang et al. [7] employed the 4th harmonic of an Er-fiber DCS system to generate UV combs spanning 385.9–387.3 nm (spectral width 3 THz), resolving 30,000 comb lines at 100-MHz spacing. To date, this is the highest number of spectrally-resolved UV comb lines achieved.

Here we present our results on producing ultrabroadband UV combs containing up to one million comb lines that are frequency resolved using DCS. The front end of our setup (Fig. 1a) is a DCS system based on a new laser platform: a pair of mode-locked Cr:ZnS lasers at a center wavelength 2350 nm and repetition rate $f_{rep}$=80 MHz that feature low phase and intensity noise, high mutual coherence, and referencing to a Rb clock (see e.g. [8]). Up to the 7th harmonic was generated in multi-grating periodically poled lithium niobate (PPLN) crystals, via cascaded $\chi^{(2)}$ process. The 0.5-mm-long PPLN had three sections with the poling periods designed for producing the 2nd, 3rd, and 4th harmonics from the 2350-nm input. The harmonics higher than the 4th were generated parasitically and, surprisingly, had enough power for DCS interferograms to be detected in real time. The average power was distributed between the generated combs (Fig. 1b) in the following way: 3W–300mW–100mW–10 mW–3.5 mW–1 mW–0.1 mW for harmonics H1 (pump laser) to H7, correspondingly.

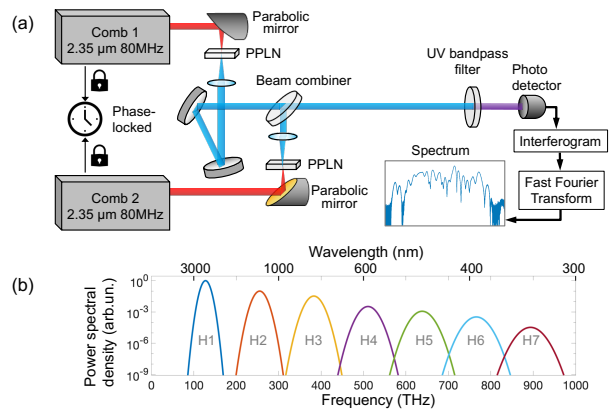

Fig. 1. (a) Schematic of the DCS setup. (b) Harmonics spectra based on the measured laser spectrum (H1) and simulations (H2–H7) under the assumption of perturbative phase-matched regime.

The UV DCS was performed by combining the two UV combs using a 50/50 beam combiner and an avalanche Si detector. For all the generated harmonics, we were able to observe comb-mode resolved spectra. Fig. 2a shows the log-scale spectrum of H6 acquired with the repetition frequency offset of $\Delta f_{rep}$=25.8 Hz and coherent averaging of 6x10^5 single interferograms (IGMs) at τ=387 min averaging time and no apodization (the averaged IGM centerburst is shown in Fig. 2b). The H6 wavelength span is 372.5-410.3 nm (spectral width 74 THz) at -30dB level (Fig. 2a), corresponding to 928,000 comb lines, while at the noise level (-40dB) the bandwidth is close to 80 THz with one million comb lines. The signal-to-noise ratio for this spectrum was SNR =588. For the number of modes M=928,000, this results in the DCS figure of merit (FOM), defined in [2], of M× SNR/$\sqrt{\tau}$=3.6×10$^6$ Hz$^{1/2}$. To reveal the comb structure, we coherently averaged four consecutive IGMs. Fig. 2c shows a portion of the related comb-line resolved spectrum.

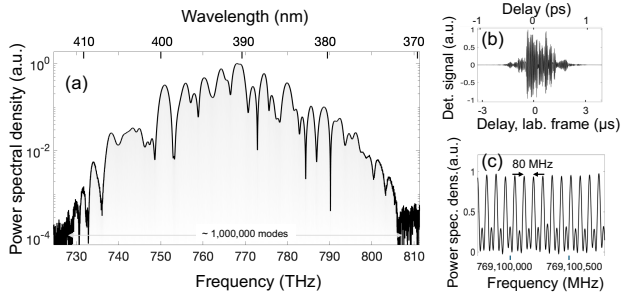

Fig. 2. (a) Spectrum of H6 containing close to a million comb lines. (b) The IGM centerburst. (c) Portion of the comb-line-resolved spectrum..

The log-scale spectrum of H7 is shown in Fig. 3a. It is derived from 3x10$^5$ coherently averaged single IGMs (acquisition time 80 min). The span is 325.4–341.8 nm (spectral width 44 THz) at -30dB level, corresponding to 550,000 comb lines and FOM=7.1×10$^5$ Hz$^{1/2}$. The IGM centerburst is shown in Fig. 3b, and the comb-resolved spectrum - in Fig. 3c. The 80-MHz comb-line spacing in this range corresponds to the wavelength resolution Δλ=30fm and the resolving power λ/Δλ=1.1x10$^7$. The noticeable intensity variations in the H7 spectrum are likely caused by destructive interferences resulting from parasitic phase-matching

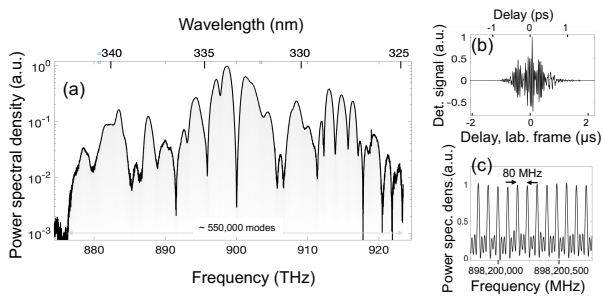

Fig. 3. (a) Spectrum of H7. (b) The corresponding IGM centerburst. (c) Portion of the comb-line-resolved spectrum.

As a spectroscopy demonstration, we measured the reflection spectrum of a volume Bragg grating (BragGrate™ Mirror from IPG/OptiGrate) with an ultra-narrow reflection band at 406 nm, tuned to 390 nm (the center of H6 spectrum) by tilting it by 22° from the normal incidence. Figure 4a shows the log-scale reflection spectrum, with the incoming spectrum shown in the background for reference. Here we used Δ$f_{rep}$= 3.05 kHz and coherently averaged a single IGM for 1s (its centerburst is shown in Fig. 4b). Fig. 4c shows an expanded view of the spectrum that has a width of 200 GHz (2,500 comb lines) at -10 dB level. Fig. 4d shows a portion of the comb-resolved spectrum with the comb-line finesse of 3,000; this spectrum was obtained by recording a 1-s-long stream of consecutive IGMs with no phase correction applied. A sharp increase in the SNR and data acquisition speed occurs here due to a decreased number of comb lines, in full agreement with [2]. In fact, this 200-GHz band can be selected anywhere within the entire H6 spectrum – by simply varying the incidence angle on the volume grating within 0…35°.

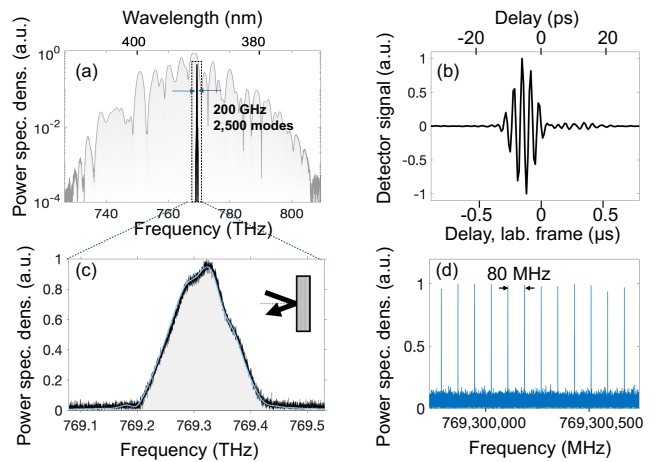

Fig. 4. (a) Reflected spectrum from the volume Bragg grating (black curve); (b) The IGM centerburst. (c) Expanded spectrum for 1s averaging (black curve) and 246 s averaging (blue curve). (d) Portion of the comb-line-resolved spectrum obtained from a 1-s stream of data.

To our knowledge, this is the first demonstration of an ultrabroadband DCS in the UV range with the capability of concurrently resolving up to a million comb lines with a single detector, all referenced to a Rb atomic clock. Although the PPLN transmission limits the UV region to >325 nm, high harmonic generation in bulk crystals such as ZnO offers a promising approach to reaching deeper UV regions, potentially down to λ≈100 nm [9]. This may pave the way toward high resolution *nuclear* laser spectroscopy, e. g. facilitating the development of a nuclear clock based on the Th-229 isotope transition near 148 nm [10].

**Funding.** DOE, Grant No. B&R #KA2601020; AFOSR, Grant No. FA9550-23-1-0126.

**Acknowledgments.** We thank Vadim Smirnov (IPG Photonics Corp./OptiGrate,) for providing volume Bragg grating samples.

**Disclosures.** The authors declare no conflicts of interest..

**Data availability.** Spectral data are available upon reasonable request.